\documentclass[twocolumn,preprintnumbers,amsmath,amssymb,prl,floatfix]{revtex4-1}

\usepackage{epsfig,psfrag}
\usepackage{dcolumn}
\usepackage{bm}
\usepackage{graphicx}
\usepackage{color}

\begin{document}

\title{Fluctuation relations and rare realizations of transport observables}

\author{Alexander Altland,$^1$ 
Alessandro De Martino,$^1$ Reinhold Egger,$^2$ and Boris Narozhny$^{1,3}$}

\affiliation{
$^1$Institut f\"ur Theoretische Physik, Universit\"at zu K\"oln,
D-50973 K\"oln, Germany\\ $^2$Institut f\"ur Theoretische Physik, 
Heinrich-Heine-Universit\"at, D-40225 D\"usseldorf, Germany
\\ $^3$Institut f\"ur Theorie der Kondensierten Materie,
Universit\"at Karlsruhe, D-76128 Karlsruhe}

\date{\today}

\begin{abstract}
 Fluctuation relations establish rigorous identities for
 the nonequilibrium averages of observables.  Starting from a general
 transport master equation with time-dependent rates, we employ the
 stochastic path integral approach to study statistical fluctuations
 around such averages.  We show how under nonequilibrium conditions,
 rare realizations of transport observables are crucial and imply
 massive fluctuations that may completely mask such
 identities. Quantitative estimates for these fluctuations are
 provided.  We illustrate our results on the paradigmatic example of
 a mesoscopic RC circuit.
\end{abstract}

\maketitle

In the past decade, the concept of fluctuation relations has become a
powerful new paradigm in statistical physics.  Generalizing the
celebrated fluctuation-dissipation theorem, fluctuation relations
establish connections between the nonequilibrium stochastic
fluctuations of a system and its dissipative properties. The ensuing
theoretical and experimental perspectives have sparked a wave of
research activity \cite{ritort,schutz,sevick,marconi,esposito}.

Fluctuation relations generally relate to observables carrying
thermodynamic significance, such as work, heat, entropy, or
currents. By way of example, consider the Jarzynski relation
\cite{jar}
\begin{equation}\label{jarrel}
\left\langle e^{-\beta W}\right\rangle = e^{-\beta \Delta F},
\end{equation}
where $W$ is the work done on a system during a nonequilibrium process
in which driving forces act according to some prescribed 'protocol'. 
The averaging $\langle \dots \rangle$ is over all realizations of the process,
$\Delta F$ is the free energy difference between equilibrium states 
with forces fixed at the initial and final value, and 
$\beta=T^{-1}$ is the inverse temperature ($k_B=e=1$ throughout). 
Relations ('theorems') of this type are usually proven with
an incentive to establish rigorous bounds on 
thermodynamic quantities.  

However, considerably less efforts~\cite{grosberg,morriss} have 
been put into quantitatively exploring the fluctuations of observables
around the constraints imposed by fluctuation relations.  
For example, for protocols
returning to the initial configuration, $\Delta F=0$, the Jarzynski
relation has the status of a sum rule, $\langle X\rangle =1$, for the
average of the statistical variable $X= e^{-\beta W}$.  Since the average
growth of entropy demands $\langle W\rangle \ge 0$, the
sum rule quantifies the presence of exceptional process realizations with
$W<0$ \cite{jar,jarRareEvents}. The exponential dependence of the
variable $X$ on $W$ implies that this variable acts as a 'filter'
whose fluctuations around the unit average contain specific
information on rare processes. The statistics of $X$ is also vital to
the applied relevance of Eq.~(\ref{jarrel}): as we 
demonstrate below, even modest departures off thermal equilibrium tend
to amplify fluctuations to the extent that the average $\langle
X\rangle =1$ cannot be resolved in practice. Conversely, observables
like $X$ may serve as diagnostic tool detecting how far away from
equilibrium the system has been driven, and what physical processes
are responsible.

In this work, we provide a theoretical approach to 
the rare event statistics of fluctuation relations
\textit{in transport}. The motivation for emphasizing
transport lies in the applied importance of 'particle current flow' to
the nonequilibrium dynamics of stochastic systems. (To give just
two examples, let us mention charge currents in electronic circuits 
and the migration of species in biological systems.) 
Conceptually, the current flow through a system acts as a
source of noise on account of the discrete nature of particle
exchange. Off equilibrium, this 'shot noise', rather than thermal noise, 
is often the dominant source of stochasticity, 
and a consistent theory must account for the ensuing 
feedback cycle of current into noise and back.
To this end, we will employ Markovian transport master
equations \cite{VanKampen,nazarov,alex} as a minimal framework resolving
the statistics of individual particle transmission events.
We analyze the master equation in terms of a stochastic 
path integral \cite{kubo,alex}, an  approach tailor-made 
to the description of rare events in analytic terms. 
 In applications to 'mesoscopic' systems, 
the path integral affords an interpretation as the semiclassical 
limit of a quantum nonequilibrium Keldysh theory \cite{alex,pilgram},
thus establishing connections between classical and quantum
fluctuations.

\begin{figure}[h]
\centerline{\scalebox{0.37}{\includegraphics{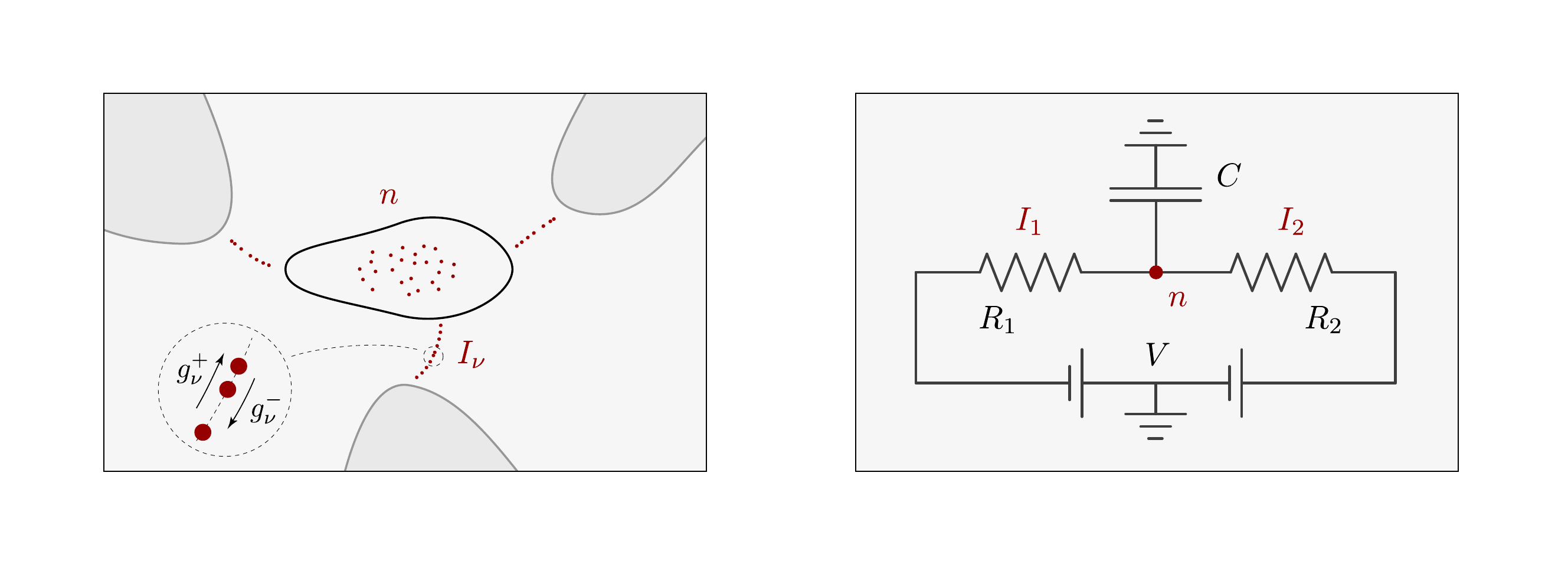}}}
\caption{ \label{fig1} 
(Color online) Left panel: Schematic illustration of a generic system 
exchanging particles with $M$ reservoirs. Right panel:
Realization in terms of an electric RC circuit.}
\end{figure}

Generally speaking, we are interested in the statistical properties of
particle currents, $I_{\nu=1,\dots ,M}$, 
flowing from some `system' to $M$ connected `reservoirs', see Fig.~\ref{fig1}. 
Assuming particle number conservation, the instantaneous number of
particles in the system $n$ is subject to a continuity equation,
$\langle \dot n \rangle  + \sum_\nu \langle I_\nu \rangle =0$. 
The system exchanges particles with the $\nu$th reservoir at
time-dependent rates $g_{\nu,t}^\pm$, see Fig.~\ref{fig1}.
We require a detailed balance condition to hold,
\begin{equation}\label{eq:1}
\frac {g^+_{\nu,t}}{ g^-_{\nu,t}} = e^{-\beta \kappa_{\nu,t}(n)},
\quad \kappa_{\nu,t}(n)=\partial_n U(n)-f_{\nu,t} ,
\end{equation}
stating that the logarithmic ratio of rates is governed 
by a \textit{cost function}  
measuring the difference in 'energies' $E_\nu(n)= U(n) -n f_\nu$
before and after a particle has entered the system through the
$\nu$th terminal,
$T\ln(g^+_\nu/g^-_\nu) =-[ E_\nu(n+1)-E_\nu(n) ]$.
Here, $f_\nu$ is a time-dependent driving force due to reservoir
$\nu$, and $U(n)$ measures the system's internal energy, where nonlinearities
in $U(n)$ describe particle interactions
and $\partial_n U(n) \equiv U(n+1)-U(n)$.
The probability $P_t(n)$ to find $n$ particles 
at time $t$ is then governed by the one-step 
master equation~\cite{VanKampen,nazarov,alex}
\begin{equation}\label{master}
\partial_t P_t(n) = -\hat H_g P_t(n),
\end{equation}
with 'Hamiltonian' $\hat H_g(n,\hat p)= \sum_{\nu,\pm}
\left(1-e^{\mp\hat p}\right) g^\pm_{\nu}(n)$, where $e^{\hat p}$
($e^{-\hat p}$) raises (lowers) $n$ by one unit.  At the initial time
$t=-\tau$, the system is assumed to be in equilibrium,
$f_{\nu,-\tau}=0$ and $P_{-\tau}(n) = \rho(n) = e^{-\beta (U(n)-F)}$
with free energy $F=-T\ln \sum_n \exp(-\beta U)$.  During the time
interval $[-\tau,\tau]$, the external forces $f_{\nu,t}$ then evolve
according to a prescribed cyclic protocol, returning to
$f_{\nu,\tau}=0$ at the final time $t=\tau$.  Equations (\ref{eq:1})
and (\ref{master}) describe
a large spectrum of transport processes in
and outside physics \cite{foot}.  Examples include charge transport in
mesoscopic devices \cite{nazarov}, molecular motors
\cite{KolomeiskyFisher}, chemical reaction networks
\cite{SchmiedlSeifert}, and evolution in biological quasispecies
models \cite{MustonenLassig}.  

The master equation (\ref{master}) deliberately emphasizes the analogy
to an imaginary-time Schr\"odinger equation with Hamiltonian $\hat
H_g$.  This formal correspondence implies that the time evolution of
$P_t(n)$ can be represented in terms of a path integral
\cite{kubo,alex}.  Applying a standard Trotter
decomposition to the unit-normalized 'partition function', $Z\equiv
\sum_n P_\tau(n)=1$, one obtains $Z=\int D(n,p) \,
e^{\int_{-\tau}^\tau dt\left(p\dot n-
   H_g(n,p)\right)}\rho(n_{-\tau})$, where the integration is over
smooth paths $\{(n_t,p_t)\}$ and the 'momentum' $p_t\in i\mathbb{R}$
conjugate to the particle number is integrated over the imaginary
axis. For technical details concerning discretization and normalization 
issues, see also Ref.~\cite{new}.
To extract information on the current profiles $\{I_{\nu,t}\}$,
we introduce sources, $Z \to Z [\chi]$, by coupling $H_g$ to 'vector
potentials' (counting fields) $\chi =\{\chi_{\nu,t}\}$,
\[
H_g(n,p)\to H_g(n,p,\chi)\equiv \sum_{\nu,\pm} 
\left(1-e^{\mp(p-i\chi_{\nu})}\right) g^\pm_{\nu}(n). 
\]
Much like in ordinary quantum mechanics, 
moments of the currents can then be generated by functional differentiation,
\begin{eqnarray}\nonumber
\langle
I_{\nu_1,t_1}I_{\nu_2,t_2}\dots \rangle&=& \frac{i\delta}{
\delta\chi_{\nu_1,t_1}}\frac{i\delta}{ \delta\chi_{\nu_2,t_2}}\dots
\Big|_{\chi=0} Z [\chi] \\ \label{eq:14}
\Leftrightarrow
Z[\chi]& =&\left\langle 
e^{-i \sum_\nu \int_{-\tau}^\tau dt\, \chi_\nu I_\nu}\right\rangle.
\end{eqnarray}
This identifies the sourceful partition function
\begin{eqnarray}\label{eq:5}
Z[\chi]&=& \int D(n,p) \, e^{-S_g[n,p,\chi]}\rho(n_{-\tau}),\\ \nonumber
S_g[n,p,\chi]&=& - \int_{-\tau}^\tau dt\left(p\dot
    n-H_g(n,p,\chi)\right)
\end{eqnarray}
as generating functional.  The current probability distribution function, 
$P[I]\equiv P[I_1,\dots,I_M]$, follows by functional integration,
\begin{equation}\label{pf}
P[I] = \int D\chi \ e^{i\sum_\nu 
\int_{-\tau}^\tau dt\ \chi_{\nu} I_{\nu} } Z[\chi].
\end{equation}
Following general principles \cite{bochkov1}, 
we now aim to relate the functional $Z$,
describing evolution governed by rates $g=\{g_{\nu,t}\}$, to the
functional $Z_b$ computed for time-reversed rates $(\hat T g)_t=
g_{-t}$, i.e., $Z_b=Z\big|_{g\leftrightarrow \hat T g}$.  Here we
have defined a time reversal operator $\hat T$  acting on 
'scalar' functions $x=(n,g,f)$ as $(\hat T x)_t=x_{-t}$, while 'vectorial' 
functions $v=(I,p,\chi)$ transform as $(\hat T v)_t=-v_{-t}$.  
Using Eq.~\eqref{eq:1}, it is straightforward to verify that the
action in Eq.~\eqref{eq:5} exhibits the invariance property
\begin{eqnarray}\nonumber
S_g[n,p,\chi] & = &  S_{\hat T g}[\hat Tn,\hat T (p-\beta 
\partial_n
U),\hat T(\chi+i\beta f)] + \\ \label{eq:6}  & + &
\beta [U(n_{\tau})-U(n_{-\tau})].
\end{eqnarray}
Inserting Eq.~(\ref{eq:6}) into Eq.~(\ref{eq:5}), and changing the
integration variables $(n,p)\to (\hat T n,\hat Tp)$, 
we obtain the symmetry relation 
$Z[\chi]  = Z_b[\hat T(\chi+i\beta f)].$
Substituting this into Eq.~\eqref{pf},  we arrive at a 
variant of the Crooks relation \cite{crooks},
\begin{equation}  \label{eq:8}
\frac{P[I]}{ P_b[\hat T I]} = e^{-\beta \int_{-\tau}^\tau dt\,
\sum_\nu f_{\nu} I_{\nu}},
\end{equation}
where $P_b$ is the current probability distribution computed for time-inverted
rates $\hat T g$.  Equation (\ref{eq:8}) was first derived in 
Ref.~\cite{bochkov1} by considering the symmetry of the operator 
$\hat H_g$ generating the system's Markovian dynamics. 
Our present derivation has the advantage that it is based  
on a path integral representation. This gives us the option 
to explore the fluctuation statistics of currents beyond the 
rigorous bound imposed by Eq.~(\ref{eq:8}). 
To elucidate this point, let us integrate Eq.~(\ref{eq:8})
over $I$ and use normalization, $\int DI P_b[\hat TI]=1$, to obtain
a variant of the Jarzynski relation
\begin{equation} \label{eq:9}
\langle X\rangle =1, \quad
X= e^{\beta \int_{-\tau}^\tau dt\, \sum_\nu f_{\nu}I_{\nu}}. 
\end{equation}
On average, the currents $\langle I_\nu\rangle \sim - \beta g_\nu f_\nu$ follow
the driving forces, which means that typically $X\sim  e^{- 2 \tau \beta^2  
\sum_\nu g_\nu f_{\nu}^2}$ is an exponentially small quantity. The average
value $\langle X\rangle =1$ is due to exceptional processes where
currents fluctuate against the driving forces \cite{jarRareEvents}. 
In this sense, $X$ acts as a selective observable filtering rare events.
Note that the exponent in $X$ has a clear physical meaning: it refers to the
work done on the system during the process.

\begin{figure}
\centering
\includegraphics[width=9cm]{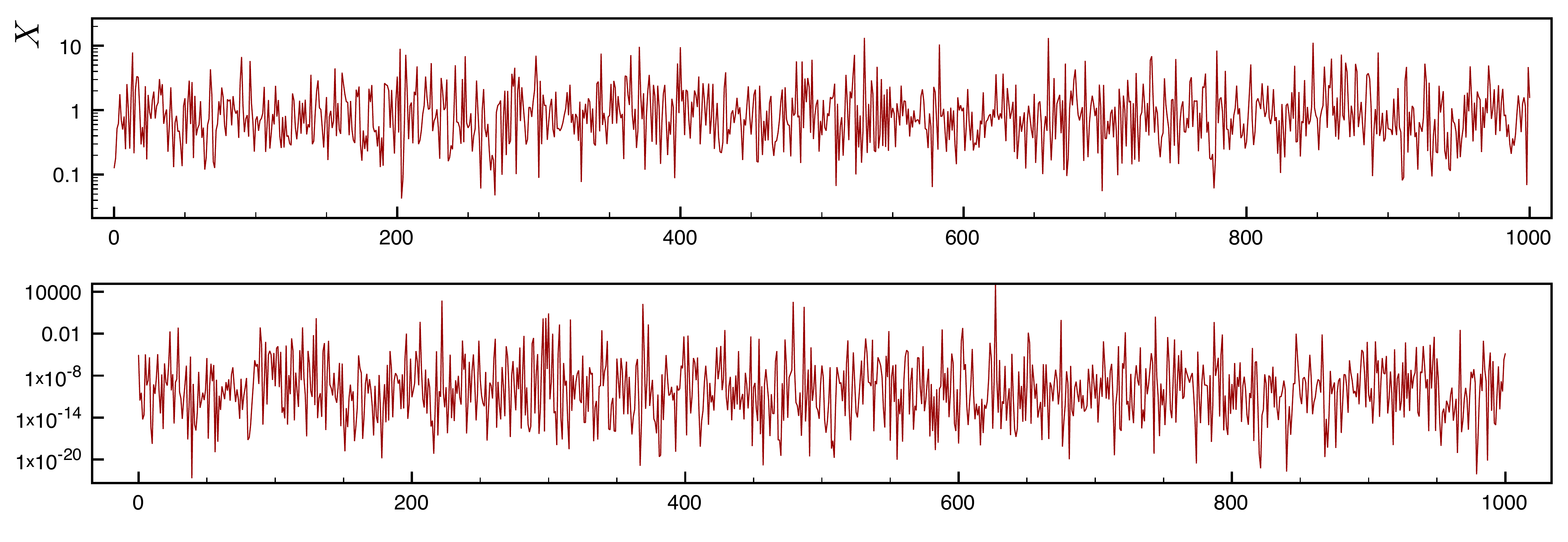}
\caption{(Color online) 
Fluctuations of the variable $X$, Eq.~(\ref{eq:9}), in the RC
  circuit over 1000 simulated process runs. For $t_V/t_T=100$ (upper
  panel), the average $\langle X\rangle=1$ remains visible. 
  However, for $t_V/t_T=1$ (lower panel),
fluctuations dominate and conceal the unit average.}
  \label{fig:X_TV}
\end{figure}

\begin{figure}[h]
\centering
\includegraphics[width=9.5cm]{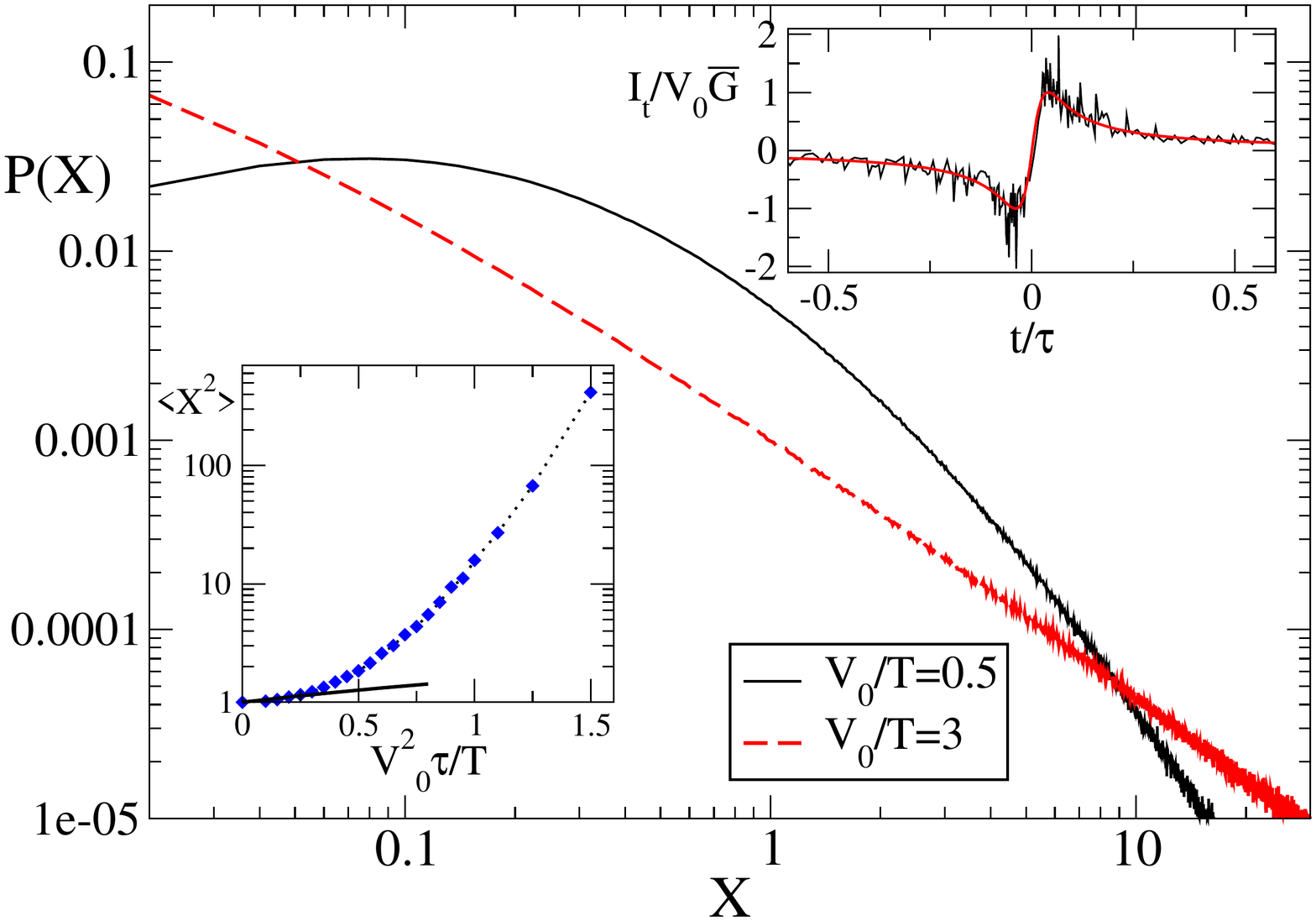}
\caption{ (Color online) Main panel: Probability distribution  of 
$X$ in Eq.~\eqref{eq:9} obtained from $2\times 10^6$ simulation
runs for the asymmetric voltage pulse 
$V_t=V_0 t_\gamma t/(t^2+t_\gamma^2)$, with $t_\gamma=0.01\tau, 
V_0\tau=250$, $\tau/t_{RC}=2\times 10^4$.  
The solid black curve ($V_0/T=0.5$) and the dashed red curve ($V_0/T=3$) 
probe the thermal and the near shot-noise regime, respectively.
Lower inset: Simulation results for $\langle X^2\rangle$ 
vs $ V_0^2\tau/T$ for $t_\gamma=0.01\tau, V_0\tau=5,
\tau/t_{RC}=2\times 10^4$.  
The straight black line is a fit to Eq.~(\ref{eq:3}) with $l=2$.
Upper inset: The black curve gives a typical current profile 
for $t_\gamma=0.04\tau$, $V_0\tau=500$, $\tau/t_{RC}=2\times 10^3$,
$\bar G=(2R)^{-1}$, and $V_0/T=10^3$. The red curve denotes the
averaged profile. }
\label{fig3}
\end{figure}

Before discussing the statistics of the variable $X$ in general
terms, let us consider the example of
a mesoscopic RC circuit biased by an external voltage, 
$f_{1/2,t}=\pm V_t/2$, see Fig.~\ref{fig1}. 
While in earlier theoretical \cite{zon1} and experimental \cite{garnier}
studies, fluctuation relations for
circuits of this type have been discussed for the 
thermal-noise dominated regime, we here consider the more general
case of noise self-generated by transport out of equilibrium.
The circuit's stochastic evolution is described by the
path integral (\ref{eq:5}) with $M=2$ and 
sequential tunneling rates \cite{nazarov},
\begin{equation}\label{raterc}
g_{\nu,t}^\pm (n) = \frac{1}{R_\nu} \frac{\pm \kappa_{\nu,t}(n)}{e^{\pm\beta 
\kappa_{\nu,t}(n)}-1},
\quad  \kappa_{\nu,t}= \partial_n U+ (-)^\nu \frac{V_t}{2},
\end{equation}
where $U(n)=\frac{e^2}{2C}(n-1/2)^2$ is the charging energy
(assuming Coulomb blockade peak conditions). 
We have performed numerical simulations of Eq.~(\ref{master})
to compare with the analytical estimates described below.
Assuming $R_1=R_2=R$,
the dynamics is characterized by a number of time scales, including the
$RC$ relaxation time $t_{RC}=RC/2$, the inverse of the mean rate at
which charges enter the central island $t_V=R/V$, the scale at
which thermal fluctuations lead to system-reservoir particle exchange,
$t_T=R/T$, and the typical scale $t_\gamma$ of variation of 
the external voltage protocol. For $t_V/t_T\gg 1$, we are in a thermal
regime where noise is of Johnson-Nyquist type and fluctuations are
comparatively benign. However, the opposite shot noise regime,
$t_V/t_T\lesssim 1$, is governed by much more aggressive
fluctuations, as illustrated by the simulation results in 
Fig.~\ref{fig:X_TV}.    Figure \ref{fig3} shows 
results for the ensuing changes in the probability distribution $P(X)$.
Comparing the two profiles, one notices the massive broadening 
of $P(X)$ upon entering the shot-noise dominated regime. The
near linearity of $P(X)$ on a double-logarithmic scale
suggests a crossover to a power-law distribution. This
indicates a divergence of all moments of $X$ except
for the first, $\langle X^{l=1}\rangle=1$. 

To make these observations more quantitative, we next consider the
moments $\langle X^{l\ge 1}\rangle$ within the path integral
approach. While details of the discussion are adjusted to the above
circuit example with $M=2$ terminals, the overall strategy is general
and can readily be adapted.  Equations (\ref{eq:14}) and (\ref{eq:9})
imply that $\langle X^l\rangle =Z[i\beta lf]$.  The introduction of
sources renders the minimum of the action $S_g[n,p,i\beta lf]$
non-vanishing. For sufficiently large observation time $2\tau$, the
action is large and stationary-phase methods apply.  We solve the
corresponding Hamilton equations, $\dot n = \partial_p H_g$ and $\dot
p = -\partial_n H_g$, under the assumption that the driving forces
vary over time scales $t_\gamma$ larger than the intrinsic relaxation
times. Again, we start from an equilibrium state and
assume a cyclic force protocol, $f_{\pm \tau}=0$.  
Under these conditions, there exist quasistationary solutions
$\dot n,\dot p\simeq 0$, and the equations become algebraic. 
The further procedure sensitively depends on whether we are in the thermal
regime defined by $\beta \tilde{f}  <1$ for all times, 
 where $\tilde{f}_t= |f_1-f_2|$, or in the off-equilibrium case 
realized otherwise.

In the \textit{thermal regime}, fluctuations around the 
averages $\langle n\rangle$ and
$\langle I_\nu\rangle$ are moderate, and the path integral can be
expanded to second order in the variables $p+ \beta l f_\nu$.  This
expansion, equivalent to the Kramers-Moyal expansion of the master
equation \cite{VanKampen}, reduces the path integral to the
Martin-Siggia-Rose functional corresponding to the Fokker-Planck
equation \cite{alex}. In this regime, we may employ the
high-temperature expansion of Eq.~(\ref{eq:1}), $g^\pm_\nu \approx
g_\nu (1\mp \frac{\beta \kappa_\nu}{2})$.  While the ensuing equations,
quadratic in $p$, afford straightforward solutions, details depend on
the $n$-dependence of the rates. We here consider the prototypical
situation of $n$-independent equilibrium rates $g_\nu$.
It is then straightforward to show that 
$p = - l \beta \sum_\nu f_\nu g_\nu/\sum_\nu g_\nu$, while
$n$ is determined implicitly by the condition $\dot n=-\langle I_1 \rangle 
- \langle I_2 \rangle =0\Rightarrow
\sum_\nu g_\nu \kappa_\nu=0$.  Substituting this configuration
into the action, we obtain
\begin{equation} \label{eq:3}
\langle X^l\rangle \sim \exp\left[ l(l-1)\beta
  \int dt \  \langle I\rangle  \tilde f  \right],
\end{equation}
where $\langle I\rangle = -\langle I_1\rangle = 
\beta \frac{g_1g_2}{g_1+g_2}  \tilde f$ 
is the average time-dependent linear-response current (we assume
$f_1>f_2$).

In the \textit{nonequilibrium regime}, however,
the shift $\chi=i \beta l f$ represents a massive (exponential) 
intrusion into the action.  The stationary phase equations are then
solved by the optimal $p$-configuration 
$p = \frac12\ln \left(\frac{\sum_\nu e^{-l\beta  f_\nu}g_\nu^+}{ \sum_\nu
e^{+l\beta f_\nu}g_\nu^-}\right),$
while the sub-exponential action dependence on $n$ is less
important. Substitution of $(n,p)$ into the action leads to 
a super-exponential scaling with the driving force \cite{fn_mismatch},
\begin{equation} \label{eq:16}
\langle X^l\rangle \sim \exp\left[ 2\int dt \langle I\rangle
 \left (-1+e^{\frac{\beta}{2}(l-1) \tilde f }\right)\right].
\end{equation}
This result shows that the moments do not diverge but become 
extraordinarily large upon entering the nonequilibrium regime. 
Once the condition $\beta \tilde{ f}_t \ll 1$ is violated, the unit 
average $\langle X\rangle=1$ 
is completely masked and the fluctuation relation (\ref{eq:9}) 
looses its practical meaning.  

Equations \eqref{eq:3} and \eqref{eq:16} exemplify how the fluctuation
statistics of $X$ changes dramatically upon departing from equilibrium
and how these changes contain telling information on the relevant 
nonequilibrium processes.  For example,
Eqs.~\eqref{eq:3} and \eqref{eq:16} have been derived for the rates
\eqref{raterc}, which in turn rely on the assumption of a uniform
temperature $T$ (generally established by external cooling.) In the
complementary case of thermal isolation, the system will heat up by
the very currents the FRs are probing. In this case, the statistics of
$X$ -- which should be straightforward to measure experimentally -- 
couples to the \textit{effective} particle distributions
building up in the system. Comparison to analytic results may then
probe the validity of models for nonequilibrium
transport and the ensuing theoretical descriptions. It is in this
sense that we believe the fluctuations of $X$  (and other
variables entering FRs) to contain 
far-reaching information beyond the rigorous FRs themselves. 

This work was supported by the SFB TR 12 of the DFG and by the
Humboldt foundation.


\begin{thebibliography}{20}
\bibitem{foot}
The generalization to reservoir-specific temperatures, 
several types of particles, or to multiple-step master equations is 
straightforward.
\bibitem{fn_mismatch} 
The derivation of Eq.~(\ref{eq:16}) neglects
terms of ${\cal O}\left([\beta\tilde f]^0\right)$. 
Therefore the exponents in Eqs.~(\ref{eq:3}) and (\ref{eq:16}) do 
not numerically match at borderline values $\beta\tilde f\approx 1$.
\end{thebibliography}
\end{document}